\documentclass[12pt]{article}

\usepackage{scicite}

\usepackage{times}

\usepackage{graphicx}% Include figure files

\newenvironment{sciabstract}{%
\begin{quote} \bf}
{\end{quote}}

\newcounter{lastnote}
\newenvironment{scilastnote}{%
\setcounter{lastnote}{\value{enumiv}}%
\addtocounter{lastnote}{+1}%
\begin{list}%
{\arabic{lastnote}.} {\setlength{\leftmargin}{.22in}}
{\setlength{\labelsep}{.5em}}} {\end{list}}

\title{Observation of the Pairing Gap\\ in a Strongly Interacting Fermi Gas}

\author
{C. Chin,$^{1}$ M. Bartenstein,$^{1}$ A. Altmeyer,$^{1}$ S.
Riedl,$^{1}$ S. Jochim,$^{1}$\\ J. Hecker Denschlag,$^{1}$ and R. Grimm$^{1,2\ast}$\\
\\
\normalsize{$^{1}$Institut f\"ur Experimentalphysik, Universit\"at Innsbruck,}\\
\normalsize{Technikerstra{\ss}e 25, 6020 Innsbruck, Austria}\\
\\
\normalsize{$^{2}$Institut f\"ur Quantenoptik und Quanteninformation,}\\
\normalsize{\"Osterreichische Akademie der Wissenschaften, 6020 Innsbruck, Austria}\\
\\
\normalsize{$^\ast$To whom correspondence should be addressed; E-mail:
rudolf.grimm@uibk.ac.at.} }

\date{}

\begin{document}

% Double-space the manuscript.

\baselineskip24pt

% Make the title.

\maketitle

% Place your abstract within the special {sciabstract} environment.

\begin{sciabstract}

We study fermionic pairing in an ultracold two-component gas of $^6$Li atoms by
observing an energy gap in the radio-frequency excitation spectra. With control
of the two-body interactions via a Feshbach resonance we demonstrate the
dependence of the pairing gap on coupling strength, temperature, and Fermi
energy. The appearance of an energy gap with moderate evaporative cooling
suggests that our full evaporation brings the strongly interacting system deep
into a superfluid state.
%the strongly
%interacting system is cooled down into a superfluid state.

\end{sciabstract}

% In setting up this template for *Science* papers, we've used both
% the \section* command and the \paragraph* command for topical
% divisions.  Which you use will of course depend on the type of paper
% you're writing.  Review Articles tend to have displayed headings, for
% which \section* is more appropriate; Research Articles, when they have
% formal topical divisions at all, tend to signal them with bold text
% that runs into the paragraph, for which \paragraph* is the right
% choice.  Either way, use the asterisk (*) modifier, as shown, to
% suppress numbering.

%\section*{Introduction}

\clearpage

The spectroscopic observation of a pairing gap in the 1950s
%\cite{firstgap1,firstgap2}
marked an important experimental breakthrough in research on superconductivity
\cite{tinkhambook}. The gap measurements provided a key to investigate the
paired nature of the particles responsible for the frictionless current in
metals at very low temperatures. The ground-breaking BCS
(Bardeen-Cooper-Schrieffer) theory, developed at about the same time, showed
that two electrons in the degenerate Fermi sea can be coupled by an effectively
attractive interaction and form a delocalized, composite particle with bosonic
character. BCS theory predicted that the gap in the low-temperature limit is
proportional to the critical temperature $T_c$ for the phase transition in
agreement with the experimental measurements. In general, the physics of
superconductivity and superfluidity goes far beyond the weak coupling limit of
BCS theory. In the limit of strong coupling, paired fermions form localized
bosons and the system can undergo Bose-Einstein condensation (BEC). The BCS
limit and the BEC limit are connected by a smooth BCS-BEC crossover, which has
been subject of great theoretical interest for more than three decades
\cite{eagles,leggett,NSR,levin}. The formation of pairs generally represents a
key ingredient of superfluidity in fermionic systems and the gap energy is a
central quantity to characterize pairing regime.

The rapid progress in experiments with ultracold degenerate Fermi gases
\cite{Cho} has opened up a unique test ground to study phenomena related to
pairing and superfluidity at densities typically a billion times below the ones
in usual condensed-matter systems. In cold-atom experiments, magnetically tuned
scattering resonances (``Feshbach resonances'') serve as a powerful tool to
control the two-body coupling strength in the gas \cite{feshbach}. Based on
such a resonance, a strongly interacting degenerate Fermi gas was recently
realized \cite{thomasScience}. A major breakthrough then followed with the
creation of Bose-Einstein condensates of molecular dimers composed of fermionic
atoms \cite{li2becinn,k2becJin,li2becMIT,li2becENS,huletBEC}, which corresponds
to the realization of a BEC-type superfluid in the strong coupling limit. By
variation of the coupling strength, subsequent experiments
\cite{markus1,jin2004,mit2004,li2becENS,kinast2004,markus2} began to explore
the crossover to a BCS-type system. This BEC-BCS crossover is closely linked to
the predicted ``resonance superfluidity''
\cite{holland,timmermans,ohashi,stajic} and a ``universal'' behavior of a Fermi
gas with resonant interactions \cite{heiselberg, universality}. The observation
of the condensation of atom pairs \cite{jin2004,mit2004} and measurements of
collective oscillations \cite{kinast2004,markus2} support the expected
superfluidity at presently attainable temperatures in Fermi gases with resonant
interactions.

Our ultracold gas of fermionic $^6$Li atoms is prepared in a balanced
spin-mixture of the two lowest sub-states $|1\rangle$ and $|2\rangle$ of the
electronic $1s^2\,2s$ ground state, employing methods of laser cooling and
trapping and subsequent evaporative cooling \cite{li2becinn}. A magnetic field
in the range between 650 to 950\,G is applied for Feshbach tuning via a broad
resonance centered at $B_0\approx830$\,G. In this high-field range,
%the nuclear spin ($I=1$) almost decouples from the electron
%spin and, for $m_s=-1/2$ electron spin projection, one obtains a
%triplet of high-field seeking states with $m_I = 1, 0, -1$. Here
%we label these states in increasing energetic order by
%$|1\rangle$, $|2\rangle$, and $|3\rangle$, respectively.
the three lowest atomic levels form a triplet of states $|1\rangle$,
$|2\rangle$, and $|3\rangle$, essentially differing by the orientation of the
nuclear spin ($m_I = 1, 0, -1$). In the resonance region with $B < B_0$, the
s-wave scattering length $a$ for collisions between atoms in states $|1\rangle$
and $|2\rangle$ is positive. Here two-body physics supports a weakly bound
molecular state with a binding energy $E_{\rm b}=\hbar^2/(ma^2)$, where $\hbar$
is Planck's constant $h$ divided by $2\pi$ and $m$ is the atomic mass.
Molecules formed in this state can undergo Bose-Einstein condensation
\cite{li2becinn,li2becMIT,li2becENS,huletBEC}. At $B=B_0$, the two-body
interaction is resonant ($a \rightarrow \pm \infty$) corresponding to a
vanishing binding energy of the molecular state. Beyond the resonance ($B>B_0$)
the scattering length is negative ($a<0$), which leads to an effective
attraction. Here, two-body physics does not support a weakly bound molecular
level and pairing can only occur due to many-body effects.

Our experimental approach \cite{li2becinn,markus1} facilitates preparation of
the quantum gas in various regimes with controlled temperature, Fermi energy,
and interaction strength. We perform evaporative cooling under conditions
\cite{details} where an essentially pure molecular BEC containing
$N=4\times10^5$ paired atoms can be created as a starting point for the present
experiments. The final laser power of the evaporation ramp allows us to vary
the temperature $T$. The Fermi energy $E_F$ (Fermi temperature $T_F = E_F/k_B$
with Boltzmann's constant $k_B$) is controlled by a recompression of the gas
performed by increasing the trap laser power after the cooling process
\cite{details}. The interaction strength is then varied by slowly changing the
magnetic field to the desired final value. The adiabatic changes applied to the
gas after evaporative cooling proceed with conserved entropy \cite{markus1}.
Lacking a reliable method to determine the temperature $T$ of a deeply
degenerate, strongly interacting Fermi gas in a direct way, we characterize the
system by the temperature $T'$ measured after an isentropic conversion into the
BEC limit \cite{details}. For a deeply degenerate Fermi gas, the true
temperature $T$ is substantially below our observable $T'$ \cite{carr,details},
but a general theory for this relation is not yet available.

Radio-frequency (RF) spectroscopy was introduced as a powerful tool to study
interaction effects in ultracold Fermi gases
\cite{debbieRF,ketterleRF,jinnature1}. Molecular binding energies were measured
for $^{40}$K atoms \cite{jinnature1}, where also the potential of the method to
observe fermionic pairing gap energies was pointed out. RF spectroscopy was
applied for $^6$Li atoms to study interaction effects up to magnetic fields of
750\,G \cite{ketterleRF}. One important observation was the absence of
mean-field shifts in the strongly interacting regime. This effect can be
attributed to the fact that, in the relevant magnetic-field range, all s-wave
scattering processes between $^6$Li atoms in the states $|1\rangle$,
$|2\rangle$ and $|3\rangle$ are simultaneously unitarity limited. This property
of $^6$Li is very favorable for RF spectroscopy as it suppresses shifts and
broadening by mean-field effects.

We drive RF transitions from state $|2\rangle$ to the empty state $|3\rangle$
at $\sim$80\,MHz, and monitor the loss of atoms in state $|2\rangle$ after weak
excitation by a 1-s RF pulse, using state-selective absorption imaging
\cite{markus1}. Our experiment is optimized to obtain a resolution of
$\sim$100\,Hz, corresponding to an intrinsic sensitivity to interaction effects
on the scale of $\sim$5\,nK, which is more than two orders of magnitude below
the typical Fermi temperatures.

We recorded RF spectra for different degrees of cooling and in various coupling
regimes (Fig.~1). We realize the molecular regime at $B=720$\,G ($a =
+120$\,nm). For the resonance region, we examined two different magnetic fields
because the precise resonance location $B_0$ is not exactly known. Our two
values $B=822$\,G \cite{mit2004} and 837\,G \cite{huletBEC,markus2} may be
considered as lower and upper bounds for $B_0$. We also studied the regime
beyond the resonance with large negative scattering length at $B=875$\,G
($a\approx-600$\,nm). Spectra taken in a ``hot'' thermal sample at
$T\approx6T_F$ ($T_F = 15\mu$K) show the narrow atomic $|2\rangle \rightarrow
|3\rangle$ transition line (upper row in Fig.~1) and serve as a frequency
reference. We present our spectra as a function of the RF offset with respect
to the bare atomic transition frequency.

Spectral signatures of pairing have been theoretically considered
\cite{zollergap1,paivi1,buechler,honew,paivi2}. A clear signature of the
pairing process is the emergence of a double-peak structure in the spectral
response as a result of the coexistence of unpaired and paired atoms. The
pair-related peak is located at a higher frequency than the unpaired-atoms
signal as energy is required for pair breaking. For understanding the spectra
both the homogeneous lineshape of the pair signal \cite{paivi1,honew} and the
inhomogeneous line broadening due to the density distribution in the harmonic
trap need to be taken into account \cite{paivi2}. As an effect of
inhomogeneity, fermionic pairing due to many-body effects takes place
predominantly in the central high-density region of the trap, and unpaired
atoms mostly populate the outer region of the trap where the density is low
\cite{bulgac,strinati,paivi2}. The spectral component corresponding to the
pairs thus shows a large inhomogeneous broadening in addition to the
homogeneous width of the pair-breaking signal. For the unpaired atoms the
homogeneous line is narrow and the effects of inhomogeneity and mean-field
shifts are negligible.
%Moreover, the simultaneous unitarity limitation of $|1\rangle$-$|2\rangle$
%interactions and $|1\rangle$-$|3\rangle$ further suppresses mean-field shifts
%of the atomic peak \cite{ketterleRF}.
These arguments explain why the RF
spectra in general show a relatively sharp peak for the unpaired atoms together
with a broader peak attributed to the pairs.

We observe clear double-peak structures already at $T'/T_F = 0.5$, which we
obtain with moderate evaporative cooling down to a laser power of $P=200\,$mW
%and $P_{\rm fin}=310\,$mW,
(middle row in Fig.~1, $T_F =3.4\mu$K). In the molecular regime ($B=720$\,G),
the sharp atomic peak is well separated from the broad dissociation signal
\cite{jinnature1}, which shows a molecular binding energy of $E_{\rm b} = h
\times 130\,{\rm kHz} = k_B \times 6.2\,\mu$K. For $B$ approaching $B_0$, the
peaks begin to overlap. In the resonance region (822 and 837\,G), we still
observe a relatively narrow atomic peak at the original position together with
a pair signal. For magnetic fields beyond the resonance, we can resolve the
double-peak structure for fields up to $\sim$900\,G.

For $T'/T_F < 0.2$, realized with a deep evaporative cooling ramp down to an
optical trap power of $P=3.8\,$mW, we observe a disappearance of the narrow
atomic peak in the RF spectra (lower row in Fig.~1, $T_F = 1.2\mu$K). This
shows that essentially all atoms are paired. In the BEC limit ($720\,$G) the
dissociation lineshape is identical to the one observed in the trap at higher
temperature and Fermi energy. Here the localized pairs are molecules with a
size much smaller than the mean interparticle spacing, and the dissociation
signal is independent of the density. In the resonance region (822 and 837\,G)
the pairing signal shows a clear dependence on density (Fermi energy), which
becomes even more pronounced beyond the resonance (875\,G). We attribute this
to the fact that the size of the pairs becomes comparable to or larger than the
interparticle spacing. In addition, the narrow width of the pair signal in this
regime (lower row in Fig.~1, $B=875$\,G) indicates a pair localization in
momentum space to well below the Fermi momentum $\hbar k_F = \sqrt{2 m E_F}$
and thus a pair size exceeding the interparticle spacing.

To quantitatively investigate the crossover from the two-body molecular regime
to the fermionic many-body regime we measure the pairing energy in a range
between 720\,G and 905\,G. The measurements were performed after deep
evaporative cooling ($T'/T_F < 0.2$) for two different Fermi temperatures
$T_F=1.2\,\mu$K and $3.6\,\mu$K (Fig.~2). As an effective pairing gap we define
$\Delta\nu$ as the frequency difference between the pair-signal maximum and the
bare atomic resonance. In the BEC limit, the effective pairing gap $\Delta\nu$
simply reflects the molecular binding energy $E_{\rm b}$ (solid line in Fig.~2)
\cite{details}. With increasing magnetic field, in the BEC-BCS crossover,
$\Delta\nu$ shows an increasing deviation from this low-density molecular limit
and smoothly evolves into a density-dependent many-body regime where
$h\Delta\nu < E_F$.

A comparison of the pairing energies at the two different Fermi energies (inset
in Fig.~2) provides further insight into the nature of the pairs. In the BEC
limit, $\Delta\nu$ is solely determined by $E_{\rm b}$ and thus does not depend
on $E_F$. In the universal regime on resonance, $E_F$ is the only energy scale
and we indeed observe the effective pairing gap $\Delta\nu$ to increase
linearly with the Fermi energy. We find a corresponding relation $h\Delta\nu
\approx 0.2\,E_F$. Beyond the resonance, where the system is expected to change
from a resonant to a BCS-type behavior, $\Delta\nu$ is found to depend more
strongly on the Fermi energy and the observed gap ratio further increases. We
interpret this in terms of the increasing BCS character of pairing, for which
an exponential dependence $h\Delta\nu/E_F \propto \exp(-\pi/2k_F|a|)$ is
expected.

In a further series of measurements (Fig.~3), we applied a controlled heating
method to study the temperature dependence of the gap in a way which allowed us
to keep all other parameters constant. After production of a pure molecular BEC
($T'< 0.2 T_F$) in the usual way, we adiabatically changed the conditions to
$B=837$\,G and $T_F = 1.2\,\mu$K. We then increased the trap laser power by a
factor of nine ($T_F$ increased to $2.5\mu$K) using exponential ramps of
different durations. For fast ramps this recompression is non-adiabatic and
increases the entropy. By variation of the ramp time, we explore a range from
our lowest temperatures up to $T'/T_F = 0.8$. The emergence of the gap with
decreasing temperature is clearly visible in the RF spectra (Fig.~3).
%The results also demonstrate the pronounced temperature dependence of the gap
%energy.
The marked increase of $\Delta\nu$ for decreasing temperature is in good
agreement with theoretical expectations for the pairing gap energy
\cite{levin}.

The situation of our experiment was theoretically analyzed for the case of
resonant two-body interaction \cite{paivi2}. The calculated RF spectra are in
agreement with our experimental results and demonstrate how a double-peak
structure emerges as the gas is cooled below $T/T_F\approx0.5$ and how the
atomic peak disappears with further decreasing temperature. In particular, the
work clarifies the role of the ``pseudo-gap'' regime \cite{levin,stajic}, in
which pairs are formed before superfluidity is reached. According to the
calculated spectra, the atomic peak disappears at temperatures well below the
critical temperature for the phase-transition to a superfluid. A recent
theoretical study of the BCS-BEC crossover at finite temperature
\cite{strinati} predicts the phase-transition to a superfluid to occur at a
temperature that on resonance is only $\sim$30\% below the point where pair
formation sets in.

We have observed fermionic pairing already after moderate evaporative cooling.
With much deeper cooling applied, the unpaired atom signal disappeared from our
spectra. This observation shows that pairing takes place even in the outer
region of the trapped gas where the density and the local Fermi energy are low.
Our results thus strongly suggest that a resonance superfluid is formed in the
central region of the trap \cite{paivi2}. Together with the observations of
resonance condensation of fermionic pairs \cite{jin2004,mit2004} and weak
damping of collective excitations \cite{kinast2004,markus2}, our observation of
the pairing gap provides a strong case for superfluidity in present experiments
on resonantly interacting Fermi gas.

\clearpage

%\bibliography{csbec}

\bibliographystyle{Science}

% Following is a new environment, {scilastnote}, that's defined in the
% preamble and that allows authors to add a reference at the end of the
% list that's not signaled in the text; such references are used in
% *Science* for acknowledgments of funding, help, etc.

\begin{scilastnote}
\item We thank P. T\"{o}rm\"{a} for a stimulating exchange of results and very
useful discussions. We thank W.\ Zwerger and H.P. B\"uchler for many
stimulating discussions. We acknowledge support by the Austrian Science Fund
(FWF) within SFB 15 (project part 15) and by the European Union in the frame of
the Cold Molecules TMR Network under Contract No.\ HPRN-CT-2002-00290. C.C.\ is
a Lise-Meitner research fellow of the FWF.
\end{scilastnote}

{\bf Supporting Online Material}

www.sciencemag.org

Materials and Methods

%Figs. S1, S2, S3

% For your review copy (i.e., the file you initially send in for
% evaluation), you can use the {figure} environment and the
% \includegraphics command to stream your figures into the text, placing
% all figures at the end.  For the final, revised manuscript for
% acceptance and production, however, Postscript or other graphics
% should not be streamed into your compiled file.  Instead, set
% captions as simple paragraphs (with a \noindent tag), setting them
% off from the rest of the text with a \clearpage as shown  below, and
% submit figures as separate files according to the Art Department's
% instructions.

\clearpage

\begin{center}
\includegraphics[width=16cm]{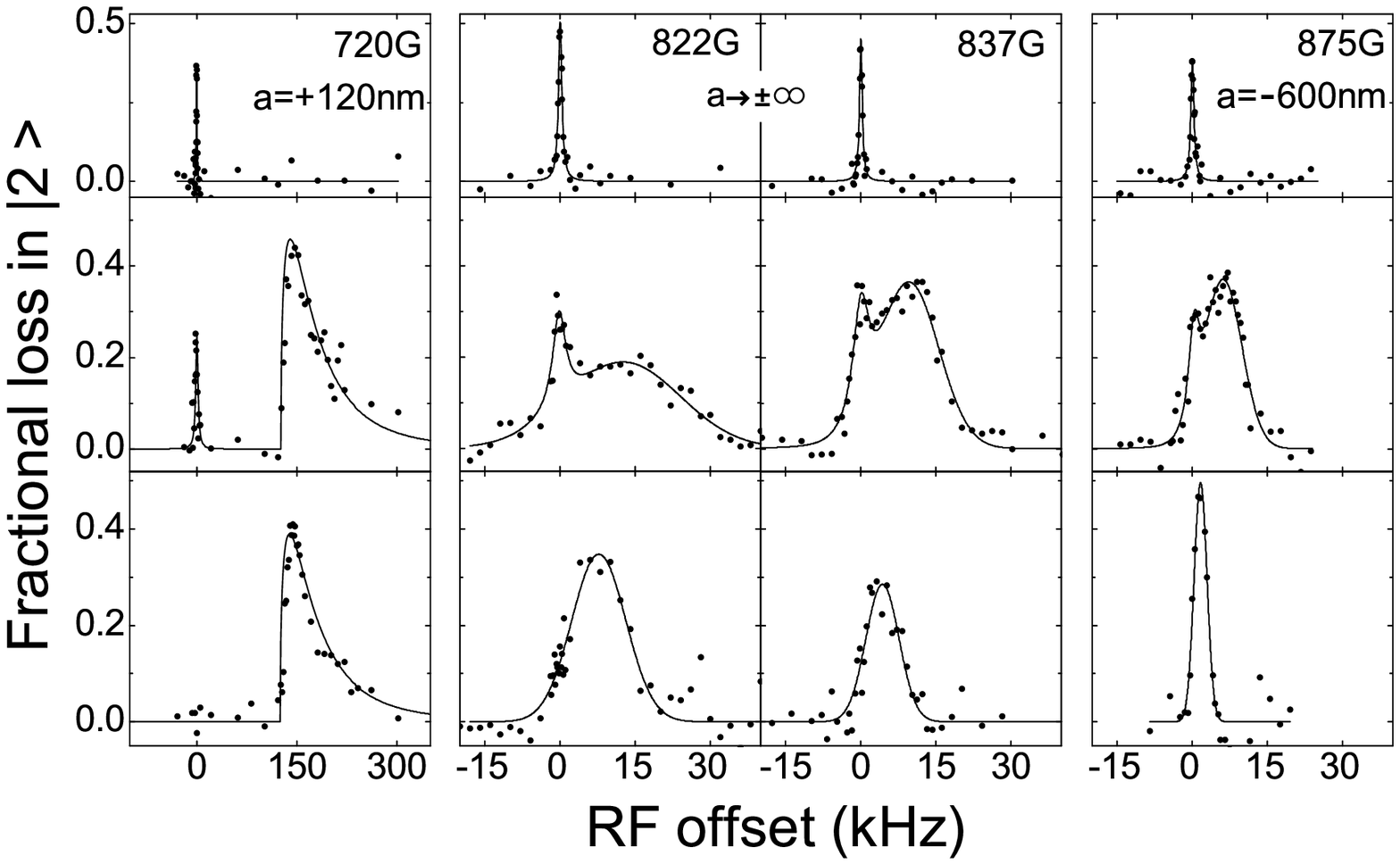}
\end{center}

\noindent {\bf Fig.~1.} RF spectra for various magnetic fields and different
degrees of evaporative cooling. The RF offset ($k_B\times 1\,\mu{\rm K} \simeq
h\times20.8\,$kHz) is given relative to the atomic transition $|2\rangle
\rightarrow |3\rangle$. The molecular limit is realized for $B=720$\,G (first
column). The resonance regime is studied for $B=822$\,G and 837\,G (second and
third column). The data at 875\,G (fourth column) explore the crossover on the
BCS side. Upper row, signals of unpaired atoms at $T'\approx 6T_F$ ($T_F
=15\,\mu$K); middle row, signals for a mixture of unpaired and paired atoms at
$T' = 0.5T_F$ ($T_F =3.4\,\mu$K); lower row, signals for paired atoms at $T' <
0.2T_F$ ($T_F =1.2\,\mu$K). Note that the true temperature $T$ of the atomic
Fermi gas is below the temperature $T'$ which we measure in the BEC limit (see
text). The solid lines are introduced to guide the eye.

\clearpage

\begin{center}
\includegraphics[width=10cm]{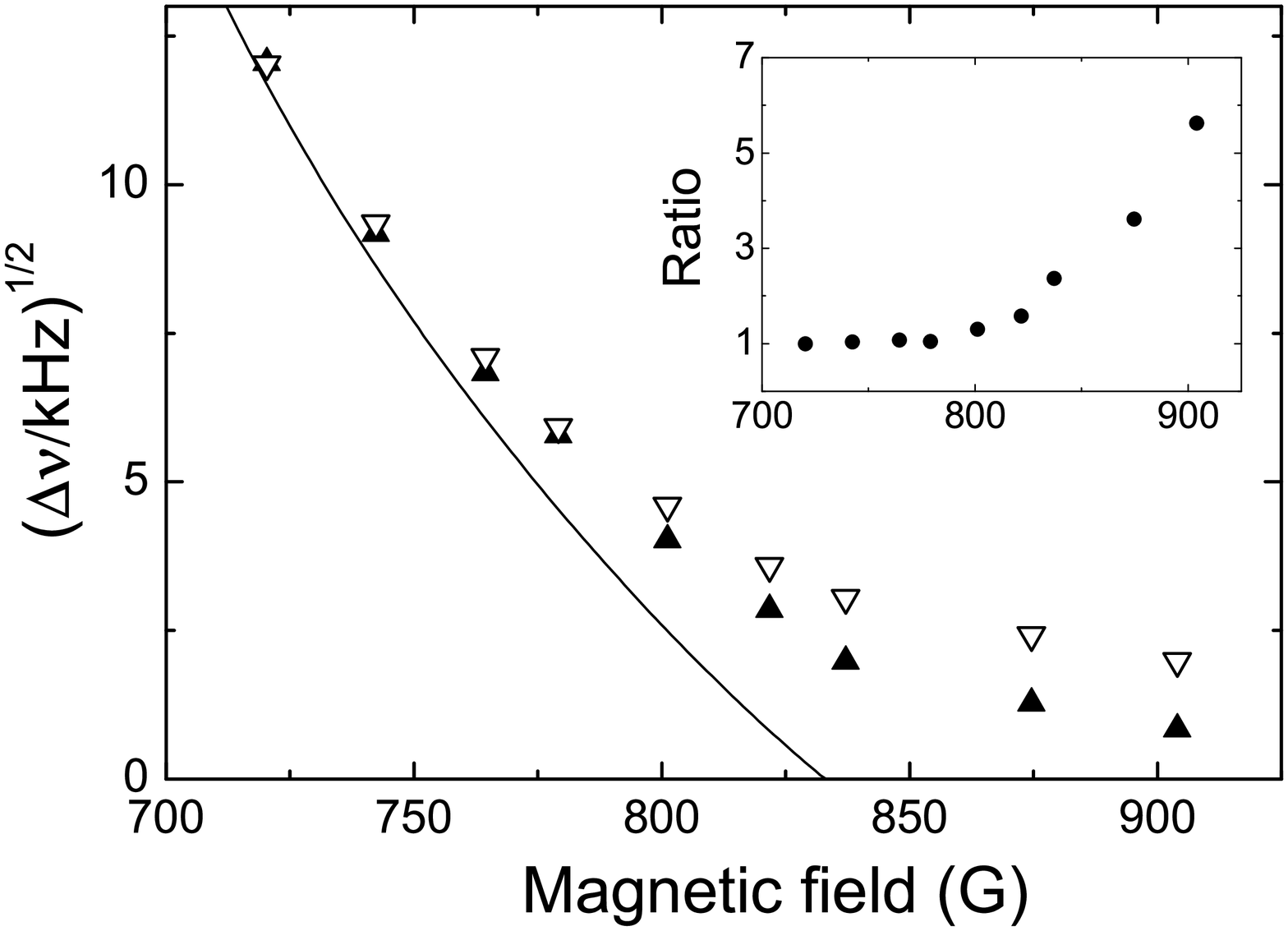}
\end{center}

\noindent {\bf Fig.~2.} Measurements of the effective pairing gap $\Delta\nu$
as a function of the magnetic field $B$ for deep evaporative cooling and two
different Fermi temperatures $T_F=1.2\mu$K (filled symbols) and $3.6\mu$K (open
symbols). The solid line shows $\Delta\nu$ for the low-density limit where it
is essentially given by the molecular binding energy \cite{details}. The inset
displays the ratio of the effective pairing gaps measured at the two different
Fermi energies.

\clearpage

\begin{center}
\includegraphics[width=7cm]{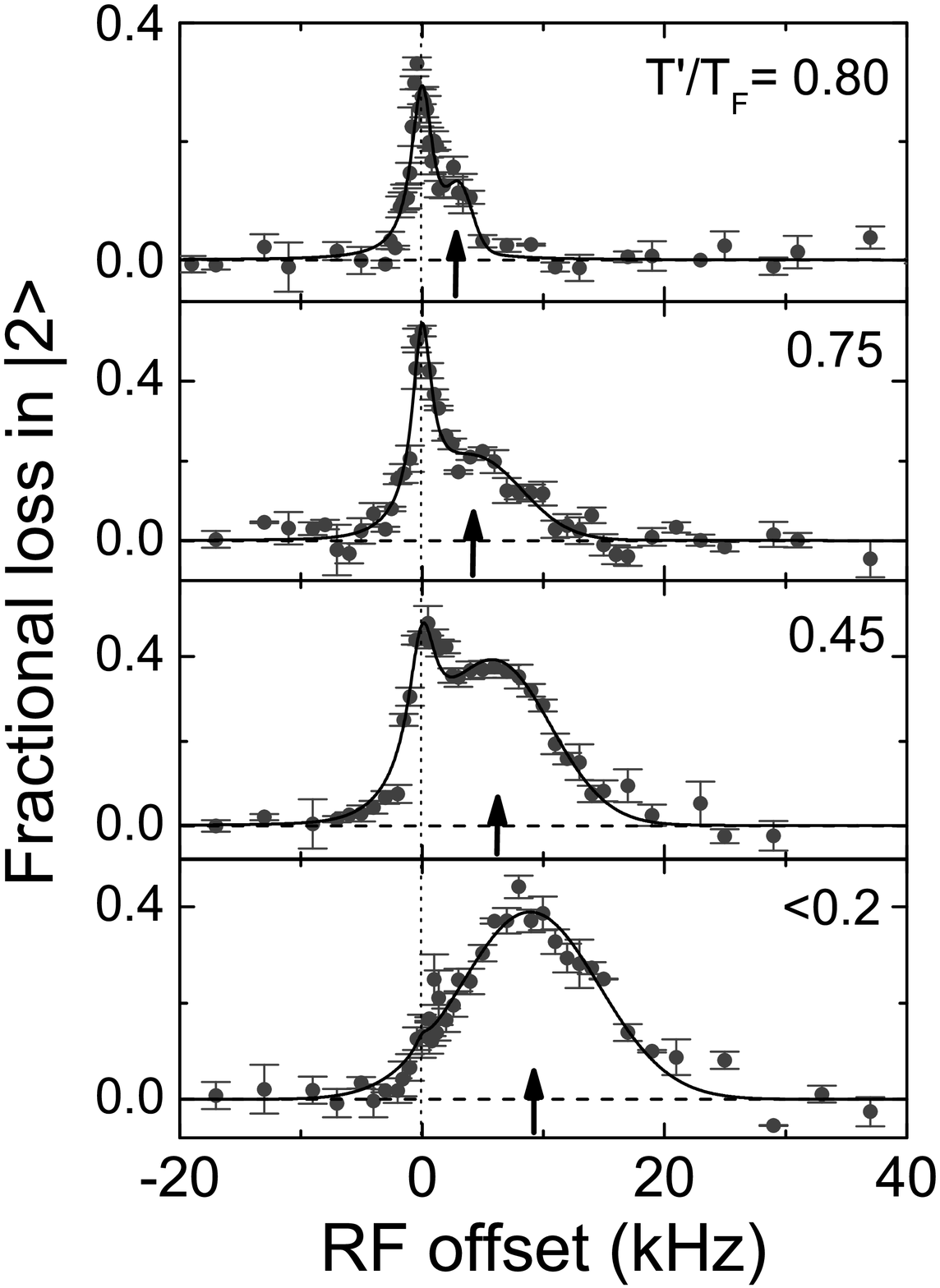}
\end{center}

\noindent {\bf Fig.~3.} RF spectra measured at $B=837$\,G and $T_F = 2.5\,\mu$K
for different temperatures $T'$ adjusted by controlled heating. The solid lines
are fits to guide the eye using a Lorentzian curve for the atom peak and a
Gaussian curve for the pair signal. The vertical dotted line marks the atomic
transition and the arrows indicate the effective pairing gap $\Delta\nu$.

\clearpage

% Double-space the manuscript.

%\baselineskip24pt

\baselineskip18pt
\begin{center}
{{\LARGE \bf Observation of the Pairing Gap in a Strongly
Interacting Fermi Gas}
\\ \vspace{0.5cm}
{\it C. Chin, M. Bartenstein, A. Altmeyer, S. Riedl, S. Jochim,
\linebreak J. Hecker Denschlag, R. Grimm} }
\end{center}
\vspace{0.1cm}

\section*{Materials and Methods}

\subsection*{Evaporative cooling, trap frequencies and Fermi energy}
Our optical dipole trap consists of a single laser beam
(wavelength 1030\,nm) focused to a waist of $25\mu$m. Evaporative
cooling is performed at a magnetic field of $B=764$\,G by
exponentially decreasing the trap laser power $P$ from initially
10.5\,W, with a fixed time constant of 460\,ms, down to a variable
final power. About $2\times10^6$ atoms are initially loaded into
the trap. After the full evaporation ramp down to a final power of
3.8\,mW, $N/2 = 2\times10^5$ molecules ($N=4\times10^5$ atoms) are
left in the trap.

The radial and axial trap frequencies $\omega_{r}$, $\omega_{z}$
are measured as a function of the laser power $P$ and described by
\begin{eqnarray}
\omega_{r}/2\pi &=& 127 \,\textrm{Hz}\times (P/\textrm{mW})^{1/2}, \nonumber \\
\nonumber \omega_{z}/2\pi &=& (601
B/\textrm{kG}+0.3P/\textrm{mW})^{1/2} \textrm{Hz}.
\end{eqnarray}
At low laser power, the axial confinement is dominated by the
curvature of the magnetic field used for Feshbach tuning.

The Fermi energy $E_F=\hbar (3\omega_r^2\omega_zN )^{1/3}$ for a
non-interacting gas is calculated from the trap frequencies and
the measured number $N$ of paired and unpaired atoms in both
internal states.

\subsection*{Thermometry}
To characterize the entropy of the gas we use the temperature $T'$
that is measured after an adiabatic and reversible (i.e.\
isentropic) conversion of the gas into the BEC limit. In this
regime better thermometry is available than in the crossover
region. We determine $T'$ at a magnetic field of 676\,G by fitting
the well-known bimodal distribution to {\it in-situ} images of the
trapped, partially condensed molecular cloud [S1]. For the full
evaporation ramp, we observe a condensate fraction of $>$$90\%$.
From this lower limit for the condensate fraction (taking into
account interaction effects) we determine an upper limit for the
temperature $T'/T_{\rm BEC} < 0.4$, where the critical temperature
$T_{\rm BEC}$ for a non-interacting molecular BEC of $M=N/2$
dimers is given by $k_B T_{\rm BEC} = \hbar(\omega^2_r\omega_z
M/1.202)^{1/3}$. Using the relation $T_{\rm BEC}=0.518\,T_F$ for a
two-component Fermi gas in a harmonic trap, we rewrite the
temperature in terms of the Fermi energy and we finally obtain
$T'<0.2\,T_F$ for our deep evaporative cooling conditions.

For the relation of the true temperature $T$ in the crossover
region to our observable $T'$, a general theory is presently not
available. However, it was theoretically shown that the isentropic
conversion of a molecular BEC with $T'/T_F\ll 1$ to a
non-interacting Fermi gas leads to a strong temperature reduction,
following a scaling $T/T_F \propto (T'/T_F)^3$ [S2]. In a similar
way, we can also expect a substantial temperature reduction when
the molecular BEC is converted into a strongly interacting gas
with resonant interactions.

\subsection*{Experimental details for Figures 1-3}
{\bf Fig.~1:} The measurements shown in the three rows are taken
for different evaporative cooling and adiabatic recompression
conditions. The relevant parameters are summarized in the
following table:
\begin{center}
\begin{tabular}{|c|c|c|c|c|}
  \hline
  % after \\: \hline or \cline{col1-col2} \cline{col3-col4} ...
   & final evap.\ power & $N$ & recompression & $T_F$ \\
  \hline
  upper row: no evaporation &10.5\,W (no ramp)& $2\times10^6$ &10.5\,W (no ramp)& $15\mu$K \\
  middle row: moderate evap. & 200\,mW & $1.0\times10^6$ & 310\,mW & $3.4\mu$K \\
  lower row: deep evaporation & 3.8\,mW & $4\times10^5$ & 34\,mW & $1.2\mu$K \\
  \hline
\end{tabular}
\end{center}

\bigskip\noindent
{\bf Fig.~2:} All measurements are taken with $N=4\times10^5$
atoms after full evaporative cooling down to a laser power of
3.8\,mW (same as in the lower row of Fig~1). The filled symbols
refer to the case of a subsequent recompression to $P=34$mW
corresponding to $T_F=1.2\mu$K, and the open symbols refer to a
recompression to $P=930$mW corresponding to $T_F=3.6\mu$K.

\bigskip\noindent
{\bf Fig.~3:} Full evaporation with same parameters as in the
lower row of Fig.~1 and in Fig.~2. After evaporation, the trap
power is increased to $P=310$mW yielding $T_F=2.5\mu$K for
$N=4\times10^5$ atoms. The recompression is performed
adiabatically (lower picture) or rapidly with different time
constants to implement controlled heating.

\bigskip\noindent {RF power:} In all measurements, we individually adjust the
RF power to obtain a maximum loss of $\sim$40\% in state
$|2\rangle$. This value is chosen as a compromise between good
signal-to-noise ratio and minimum perturbation of the system. The
RF is generally weak and applied to the sample for a long time of
1\,s to avoid broadening effects.

\subsection*{RF molecular dissociation lineshape in low-density limit}
%Radio-frequency transitions on unpaired atoms are insensitive to the external
%degrees of freedom. Doppler effects and recoil energy are negligible.

The radio-frequency $\nu_{\rm RF}$ couples the atomic states
$|2\rangle$ and $|3\rangle$. For a non-interacting atomic gas, the
transition frequency $\nu_{23}$ as a function of the magnetic
field $B$ is determined by the well-known Breit-Rabi formula. In
the magnetic-field range of interest the nuclear spin almost
decouples from the electron spin (Paschen-Back regime) and
$\nu_{23}$ is about 80\,MHz. We introduce $\delta\nu_{\rm RF} =
\nu_{\rm RF} - \nu_{23}$ as the RF offset from the atomic
resonance.

In the case of a weakly bound molecule, formed with a binding
energy $E_b$ by an atom in state $|1\rangle$ and an atom in state
$|2\rangle$, the RF can dissociate the molecule if the threshold
condition $h\delta\nu_{\rm RF}>E_b$ is fulfilled. Above threshold,
the RF-induced dissociation produces two atoms with a total
kinetic energy of $2E_k$. In the center-of-mass system, where the
molecule is at rest, energy conservation reads
%In general, the transition frequency can be shifted by interactions
%In a dense sample, the atomic transition frequency can be shifted by mean-field
%effects. However, mean-field shifts for $^6Li$ are strongly suppressed in the
%crossover regime when $k_F|a|>1$. In the range of interest 700G$<B<$900G, all
%scattering processes among atoms in state 1, 2 and 3 are unitarity limited due
%to the occurrence of Feshbach resonances in these channels. Furthermore, the
%radio-frequency transition in a fermionic system is intrinsically immune to the
%mean-field effects between the two relevant states, namely, state 2 and state 3
%in our experiment. The insensitivity to mean-field effects was studied in Ref.
%[1,2], which provide us unbiased grounds for resolving very small pairing gaps
%$\Delta\nu\ll E_F$.
\begin{eqnarray} \nonumber
h \, \delta\nu_{\rm RF}= E_b + 2E_k\,.
\end{eqnarray}

The lineshape of the continuous RF dissociation signal can be
understood in terms of the Franck-Condon overlap of the molecular
wave function with the wavefunction in the dissociation channel (a
pair of atoms in the states $|1\rangle$ and $|3\rangle$). The
wavefunction of the molecule (dimer of atoms in $|1\rangle$ and
$|2\rangle$) is essentially determined by the scattering length
$a$ (or the corresponding binding energy
$E_b\approx\hbar^2/ma^2$). The dissociation-channel wavefunction
in general depends on the kinetic energy $E_k$ and the scattering
length $a_{13}$. For $a\ll|a_{13}|$, we find that the dependence
on $a_{13}$ has negligible influence on the lineshape. However,
for $^6$Li the dissociation channel exhibits a broad Feshbach
resonance at $\sim$690\,G, which significantly affects our
dissociation lineshape. A calculation of the lineshape [S3] yields
the expression
\begin{eqnarray} \nonumber
f(E) \propto E^{-2} (E-E_b)^{1/2}(E-E_b+E')^{-1} \,
\end{eqnarray}
where $E = h\,\delta\nu_{\rm RF}$ and $E'=\hbar^2/ma_{13}^2$.

According to our definition, the effective pairing gap $h
\Delta\nu$ in the low-density molecular limit is directly given by
the maximum of the molecular dissociation signal.  From the above
lineshape $f(E)$ it is straightforward to calculate the signal
maximum
\begin{eqnarray} \nonumber
h \Delta\nu=\xi E_b\,,
\end{eqnarray}
where $\xi$ weakly depends on $E'/E_b$ and varies between the two
limits $\xi=1$ and $\xi=4/3$ for $E'\ll E_b$ and $E'\gg E_b$,
respectively.

The magnetic-field dependence of $\Delta\nu$ that we show as solid
line in Fig.~2 is calculated on the basis of the above lineshape
$f(E)$ and the most recent data for the scattering lengths $a$ and
$a_{13}$
(and the corresponding energies $E_b$ and $E'$) from the NIST group [S4].\\ \\

\large \noindent {\bf References}\normalsize
\begin{itemize}
%\item[S1.] S. Gupta et al., {\it Science} {\bf 300}, 1723 (2003).

%\item[S2.] M.W. Zwierlein, {\it Phys. Rev. Lett.} {\bf 91}, 250404 (2003).

\item[S1.] M. Bartenstein {et al.}, %, A. Altmeyer, S. Riedl, S. Jochim, C. Chin, J. Hecker Denschlag, and R. Grimm,
{\it Phys. Rev. Lett.} {\bf 92}, 203201 (2004).

\item[S2.] L. D. Carr, G. V. Shlyapnikov, Y. Castin, {\it Phys.
Rev. Lett.} {\bf 92}, 150404 (2004).

\item[S3.] C. Chin, P. Julienne, to be published.

\item[S4.] P. Julienne, A. Simoni, E. Tiesinga, private
communication.

\end{itemize}
\end{document}